\documentclass[superscriptaddress,twocolumn,pre]{revtex4}
\usepackage{amssymb, amsmath, graphicx}
\usepackage{hyperref}
\def\be{\begin{equation}}
\def\ee{\end{equation}}
\begin{document}
\title{Dynamical Persistency in River Flows}
\author{Hyun-Joo \surname{Kim}}\email{hjkim21@knue.ac.kr}
\affiliation{
Department of Physics Education, Korea National University of Education,
Chungbuk 363-791, Korea}

\received{\today}

\begin{abstract} 

The universal fractality of river networks is very well known, however understanding of the underlying
mechanisms for them is still lacking in terms of stochastic processes. By introducing
probability changing dynamically, we have described the fractal natures of river networks stochastically.
The dynamical probability depends on the drainage area at a site that is a key dynamical quantity of
the system, meanwhile the river network is developed by the probability, which induces dynamical
persistency in river flows resulting in the self-affine property shown in real river basins, although the 
process is a Markovian process with short-term memory.

\end{abstract} 

\maketitle

The fractal properties shown in a variety of patterns of nature have been attracted attentions in many fields
for several decades  and the nature of self-similarity or self-affinity of the patterns are well described by 
scaling laws  \cite{fractal}. A field in which such fractal natures have been well 
substantiated by the extensive measurements for real river basins is hydrology
\cite{review1, review2}. The structures of river network are characterized by unified scaling laws, regardless of
size, vegetation, geological features, climate, orientation of the basin, {\it etc} \cite{scaling1, scaling2}.
The Hack's law \cite{hack} is the representative law describing the fractal geology in river networks through the 
relation between the main stream length $l$ and the river basin $A$ giving as $l \sim A^h $, where $h$
is the Hack's exponent having the values in the range $0.56 - 6.0$ for real river basins. The drainage area $a$ that
can surrogates a river basin at a given site is defined as  the number of sites connected to the actual site
through drainage direction,  that is, the area of land that collects precipitations contributing to the network,
and the distribution of drainage areas $p(a, L)$ obeys a scaling form with the following scaling function \cite{review1,scaling1},
\begin{equation}
p(a,L) = a^{-\tau} f \left( \frac{a}{a_c (L) } \right),  ~~ \Big \{
\begin{array} {ll}
\lim_{x \to \infty} ~f(x) = 0 \\
\lim_{x \to 0} ~f(x)=\text{const}.
\end{array}
\label{scaling}
\end{equation}
where the characteristic area $a_c ( L) $ satisfies the relation
\be
a_c (L) \sim L^\phi
\ee
with $\phi = 1+H$ of which relation is drawn from a river basin with a longitudinal length $L$ and a perpendicular length
$L_\perp$ following the power-law behavior between them as $L_\perp \sim L^H$. 
The exponent $H$ is called the Hurst exponent and for the cases where the main stream length $l$ is linearly
proportional to the longitudinal length $L$, the scaling relations relate the exponents $h$ and $\tau$ in terms of
$H$ as $h=1/1+H$ and $\tau=(1+2H)/(1+H)$, respectively \cite{soc2}. It means that  we may concentrated on the Hurst exponent 
in studying the fractal properties of river basins. Also the Hurst exponent for real basins is in the range of 0.75-0.80 that is
greater than 1/2, which indicates there is  persistency in river flows.

 \begin{figure*}[ht]
 \includegraphics[width=\textwidth]{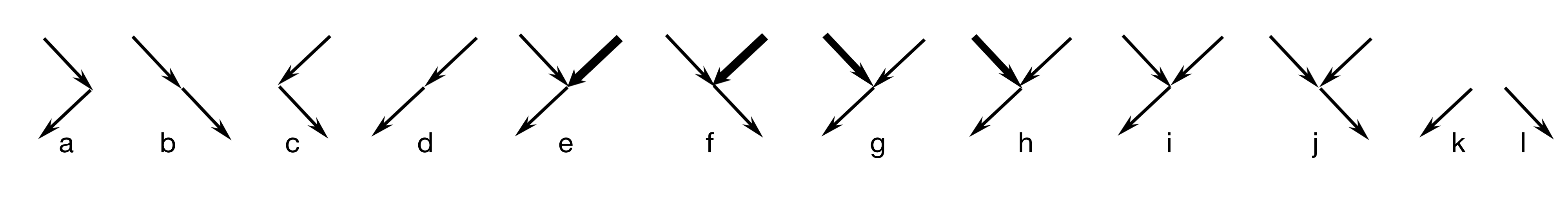}
 \caption{\label{rule} 
 All possible water-flow cases being able to appear by this stochastic model. The bolder line indicates
 that the drainage area at that site is larger than that of the other site represented with normal lines.
 When the position of a center is  $(x,y)$, the dynamical probabilities for several cases are as follows :
 for a,  $\alpha[a]=a^{-\delta} (x-1,y-1)$, for e, $\alpha[a]=1-(a(x+1,y-1)-a(x-1,y-1))^{-\delta}$, and
 for from i  to l, $\alpha[a]=1/2$.
 }
 \end{figure*}

Elucidating the underlying mechanisms  for the fractal natures of real river basins is very interesting and important, 
and several models have been suggested with quite different origins to do so.
First of all, the optimal channel network (OCN) \cite{ocn1, ocn2, ocn3} developed river networks 
based on the integrating principle of minimizing the total energy dissipation defined as $E= \sum _i a_i ^{1/2}$ in the
system. Second, river networks developed  on the landforms sculpted mostly by local erosions 
\cite{e1,e2,e3,e4,e5} or formed by avalanches evoking a long-range slope-slope correlation \cite{rn},
have been introduced.  
Third, the evolution of river network has been considered by the framework of the self-organized criticality \cite{soc}
which describes a threshold dynamics resulting in the fractal natures in a process \cite{soc1,soc2}.
Finally, in  the Scheidegger model \cite{scheidegger} where  the drainage direction of waterflows is determined
only probabilistically without taking account of any detailed contexts of the basin,  water flows are regarded
as a stochastic process.
Models based on OCN, evolution of landforms, and SOC got at the exponent values measured in real  river basins, 
however, the stochastic model for water flows is mapped into the well-known random walk model with $H=1/2$ 
and thus does not well explain the pattern of river basins showing the persistency in flows. 
For stochastic diffusive phenomena several representative models such as the fractional Brownian motion (fBM)  \cite{fbm},
the continuous time random walks (CTRW)  \cite{ctrw}, and L\'{e}vy walk model \cite{levir,leviwalk}
have well described the persistent property, however, they do not directly explain how is water-flows persistent
in river basins.

Recently, the memory effects of previous history have been introduced as the main origin inducing 
the persistency in a stochastic process \cite{mem1,mem2,mem3,mem4,kim2014}. Also we have introduced the 
nonstationary Markovian replication process (NMRP)  that can induce the persistency by taking into consideration 
only the shortest-term memory \cite{nmrp}.
In NMRP, the direction of a next step is determined by replicating that of  the just previous step with a replication probability
which changes at each time. For the constant replication probability, it becomes to the persistent random-walk model \cite{prw}
and in continuum limit, the telegrapher equation \cite{tele1,tele2,tele3} which initially introduces the wave equation property and is
related to the ballistic motion of diffusive particles, but it reduces asymptotically to normal diffusive behavior.
Unlike the persistent random-walk model, the time-varying replication probability of NMRP controls the tendency 
to move in the same direction over time, which reflects adaptations to act in response to changes of environment and
give rise to the persistency (or antipersistency ) despite Markovianity of a process.
Likewise, for water flows in river basin, direction of water flowing into a site can influence next
direction of outflow from the site at each moment differently. Therefore we can consider the origin of the persistency
of river flows from the viewpoint of NMRP.
Furthermore, for the real complex contexts it may be more plausible for the probability to depend on a
dynamical quantities  mainly characterizing a system rather than explicitly on time. 
In this study, we have attempted to generalize the NMRP to the problem of river network.
As the drainage area at each site is a key dynamical quantity characterizing river networks, we have
taken it into consideration as surrogating time and have developed the river basin that shows the
fractal natures of real river basins.

 The details of rules of the generalized NMRP model are as follows:
Consider a lattice of size $L_\perp \times L$ in which the position of a site is denoted as $(x,y)$ with 
$1 \leq x \leq L_\perp$ and $1 \leq y \leq L$. Initially, a water drop is allocated at each site, that is, $a(x,y)=1$ reflecting
uniform rainfall and $\sigma(x,y)$ is a random variable that determines the direction of water flows
at site $(x,y)$ and initially is assigned with 1 or -1 at random.
When $\sigma(x,y)=1$ water at site $(x,y)$ flows down into the right site $(x+1, y+1)$ and when $\sigma(x,y)=-1$
the water flows into the left site $(x-1, y+1)$. Here, flow from a site $(x, y)$ into $(x, y+1)$ is not allowed for just 
simplicity, which does not affect the results. For $y \geq 2$, $\sigma(x,y)$ is determined by the probability
$\alpha[a]$ as follows,
\be
\sigma(x,y) = \left \{
\begin{array}{ll}
\sigma_0,  & \quad \text{with probability $\alpha[a]$}\\
-\sigma_0,  & \quad \text{with probability $1-\alpha[a]$.}
\end{array} \right.
\label{erule}
\ee
Here if $w_+  a(x-1,y-1) >w_-  a(x+1,y-1)$,  then $\sigma_0 =\sigma(x-1,y-1)$, if  
$w_+  a(x-1,y-1) <w_-  a(x+1,y-1)$, then $\sigma_0 =\sigma(x-1,y-1)$
and if the two have the same value then $\sigma_0$ is assigned the value of 1 or -1 which is randomly chosen,
and if water flows from a left (right) site into the site $(x,y)$, $w_+ (w_-) =1 (-1)$, and if not, $w_{\pm}=0$,. 
That is, if water from a neighboring site flows into the actual site, $\sigma_0$ at that site is the same that of the
neighboring site, while if water flows from both of the two neighboring sites into that site, $\sigma_0$ takes 
the value of $\sigma$ of the site with a larger drainage area.

 \begin{figure}[ht]
 \includegraphics[width=9cm]{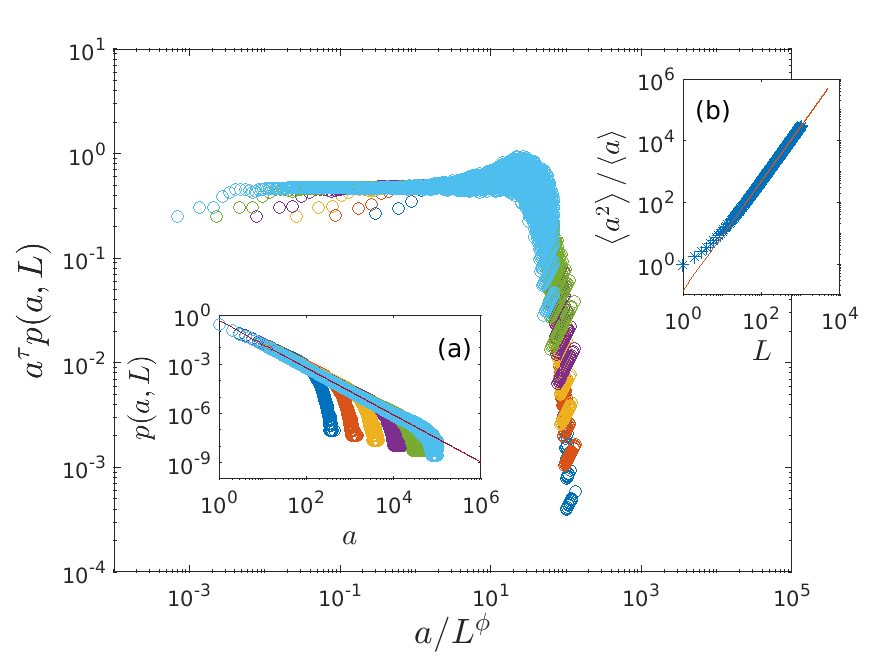}
 \caption{\label{pa}
 The plot of the scaling function of $p(a,L)$ for various values of $L=32, 64, 128, 256, 512$, and $1024$ 
 with $\delta=0.25$.
 The data are very well collapsed with $\tau = 1.44$ and $\phi = 1.78$. The inset (a) shows the plot of $p(a)$ and the
 straight guide line represents $\tau \approx 1.43$. In the inset (b), $\phi \approx 1.77$ has been obtained
 from the relation, $A= \left< a^2 \right> / \left<a \right> \sim L^\phi$. 
 }
 \end{figure}
 \begin{figure*}[ht]
 \includegraphics[width=\textwidth]{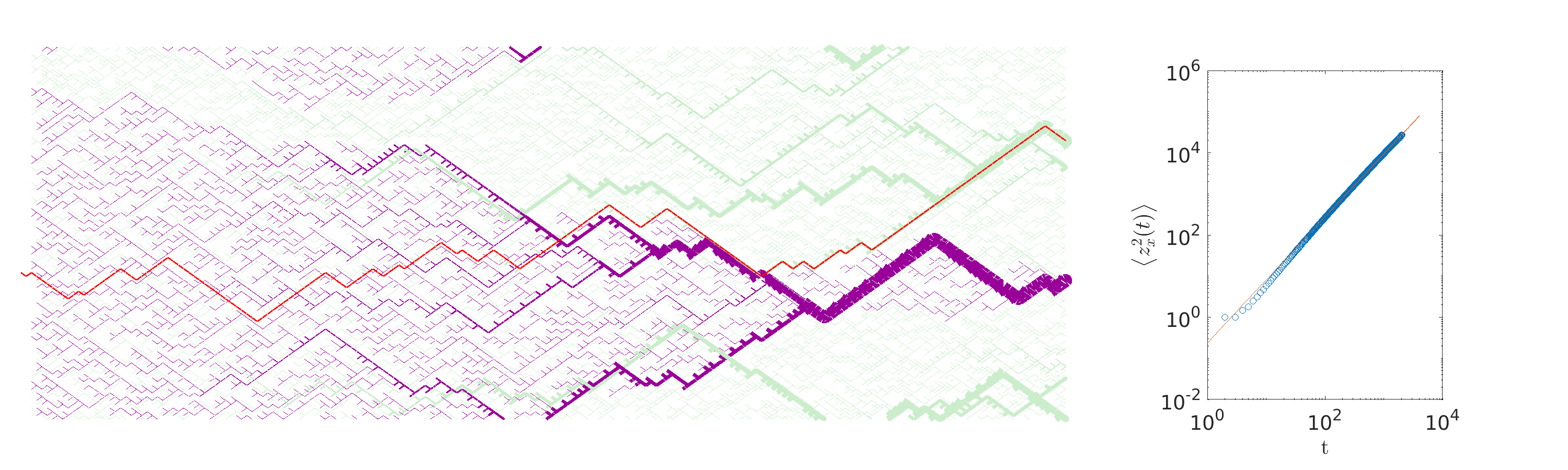}
 \caption{\label{basin} 
The river network drawn by this stochastic model. The links sharing a single largest sink are 
depicted as purple lines.
The red line is a sample of the entire path of water flow starting from a initial site, and the MSD measured 
from the paths is shown in the right and  the slope of the straight guide line represents $H \approx0.77$.
 }
 \end{figure*}

The definition of the probability $\alpha[a]$ is as follows,
\be
\alpha [a] = \left \{
\begin{array}{ll}
1 - \Delta^{-\delta},  & \quad \text{if $\Delta \neq 0$}\\
1/ 2,  & \quad \text{if $\Delta = 0$.}
\end{array} \right.
\label{alpha}
\ee
where $\delta >0$ and $\Delta = | w_+ a(x-1,y-1) - w_- a(x+1, y-1)|$.
The probability $\alpha[a]$ are not explicitly dependent on time and thus not predetermined unlike NMRP.
Rather, it is dynamically generated according to the key dynamical quantity, i.e., the drainage area
every time as the process evolves. Thus we call it the dynamical replication probability.
That means the water flows along the same direction in $x$
of previous flow $\sigma_0$ with the probability $\alpha$ that becomes larger as $\Delta$ is larger.
$\Delta$ can be regarded as a momentum of water at that site if it is regarded as its velocity is uniform 
and thus the larger the momentum is over time, the 
stronger the tendency to keep the direction of the flow becomes. 
In other words, it brings about the dynamical persistency in the sense that the persistency is 
determined by the drainage area changing dynamically at each moment. 
Figure \ref{rule} shows the schematic pictures for all cases water can flow from a site into downward sites
according to the previous movement.

To find out if this stochastic model well works, we measured the distribution of drainage areas $p(a, L)$
for several longitudinal size $L$ with $\delta=0.25$.
The inset (a) of the Fig. \ref{pa} shows $p(a) \sim a^{-\tau}$ with $\tau \approx 1.43$ and the main panel shows
the data collapse of $p(a,L)$ with $\tau = 1.44$ and $\phi = 1.78$. Also to
independently measure the exponent $\phi$, we used an alternative definition of the characteristic 
area $a_c = \left< a^2 \right> / \left<a \right> \sim L^\phi$ of
which the scaling relation is able to be obtained from the Eq. \ref{scaling},
The inset (b) shows $\phi \approx 1.77$ that is excellent agreement with that obtained 
from the data collapse.
Thus the values of the exponents are excellent agreement with those of real river basins, which
indicates that this stochastic model very well develops real river basins.
In the vast majority of stochastic models, the probability is either constant, or some fixed function of parameters 
and is not a dynamical variable by itself. Promoting the probability itself to be a dynamical variable in our model allows 
us to capture the universal features of the river network without having to take into consideration any external parameters.

Figure \ref{basin} shows a pattern of a river network drawn by this model. In the background, all paths
of flowing water from at each site into sinks in the bottom line has been drawn and the purple lines on the
background pattern represent the paths connecting to the largest sink among them, which looks like 
a real river network. Meanwhile, the blue path that is a sample line from 
a top site to the bottom site,  can be treated as a one dimensional stochastic process. 
To further clarify the properties from the view of random walk in one dimension, we consider 
$z_{x} (t)$ that is a position of a water starting from a $x$ site 
initially, at time $t$ that surrogates a position $y$. The evolution of it is given as
\be
z_{x} (t+1)=z_{x}(t)+\sigma_{x}(t)
\ee  
where $\sigma_{x}(t) = \sigma(x(y),y)$ and $x(y)$ is a $x$ position of water starting from $(x, 1)$ site at $y$.
The value of mean-squared displacement (MSD) obtained by averaging these paths is $H \approx 0.77$ 
for $\delta=0.25$ as shown in the right of 
the Fig.\ref{basin}, which is excellent agreement with that of obtained from the relation of $\phi = 1+H$.
It means that the deviation of the stochastic process composed by these one dimensional paths
represents the width of the river basin.

\begin{figure}
  \includegraphics[width=8cm]{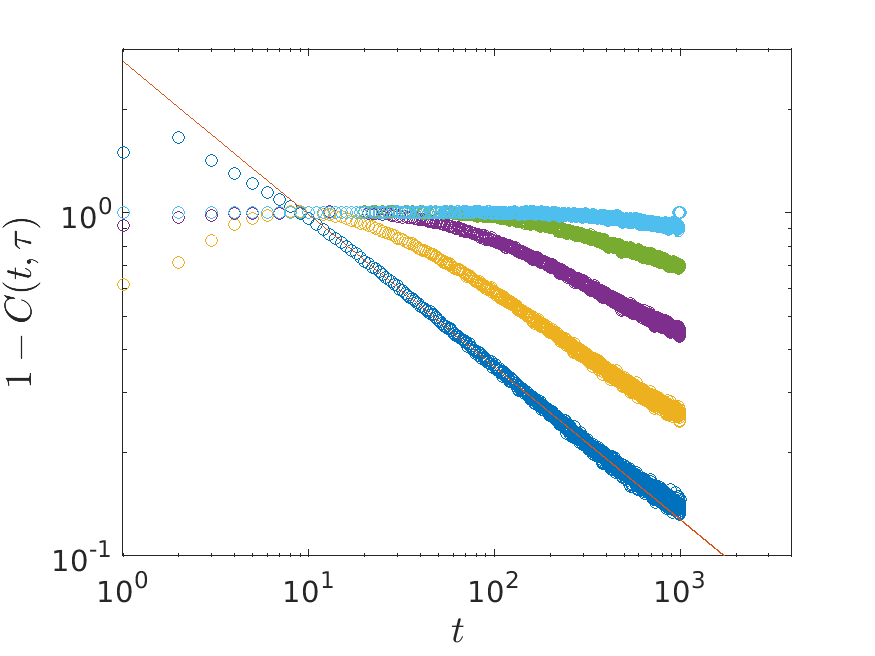}
  \caption{The plot of $1-C(t, \tau)$ as a function of time $t$ for various
  $\tau = 1, 2, 4, 8, $ and $16$ from the bottom to the top. The guide line for $\tau=1$ represents
  the value of $m$ that is the parameter controlling
  the replication probability of the NMRP is 0.45 and the corresponding $H$ has a value of 0.72 which
  is different from that obtained for the river network. When $\tau$ become larger, the slope is
  smaller, which is also different results from that of the NMRP where the slope for large time
  is same irregardless of $\tau$.}
  \label{cv}
\end{figure}

To find out correlations between the steps we considered the step-step correlation function defined as 
\be
C(t, \tau) = \left < \sigma(t)\sigma(t+\tau)  \right>
\ee
and have numerically found that $C(t,1) \sim 1- t^{-0.45}$ as shown in the Fig. \ref{cv}.
In order to compare this to the NMRP in which next step is determined as follows\cite{nmrp}
\be
\sigma({t+1}) = \left \{
\begin{array}{ll}
\sigma(t),  & \quad \text{with probability $1-\alpha(t)$}\\
-\sigma(t),  & \quad \text{with probability $\alpha(t)$,}
\end{array} \right.
\label{erule}
\ee
and for $\alpha(t)$ not close to 1, the MSD is given asymptotically as \cite{nmrp}
\be
\langle z^2 (t) \rangle \sim \int_{}^{t} {\alpha(s) \over {1-\alpha(s)}  } ds,
\label{msd}
\ee
we have calculated the step-step correlation function for $\sigma(t)$ as 
\be
C(t, \tau) = \Pi_{s=t+1}^{t+\tau} [ 2\alpha(s) - 1 ].
\ee
Because $\alpha[\Delta]$ is power functional of $\Delta$ that linearly depends on time,
considering $\alpha(t) =1- t^{-m} (m>0)$, the correlation function becomes $C(t, \tau) = \Pi_{s=t+1}^{t+\tau} (1-2/t^m )$ 
and for enough large time, $t \gg \tau$,
\be
C(t, \tau) \approx \left ( 1- {2 \over t^m} \right)^\tau \approx 1- {2\tau \over t^m}
\ee
and the relation between the Hurst exponent and the parameter $\delta$ is given by $2H = m+1$ for large $t$  \cite{kim2014}.
\begin{figure}
  \includegraphics[width=8cm]{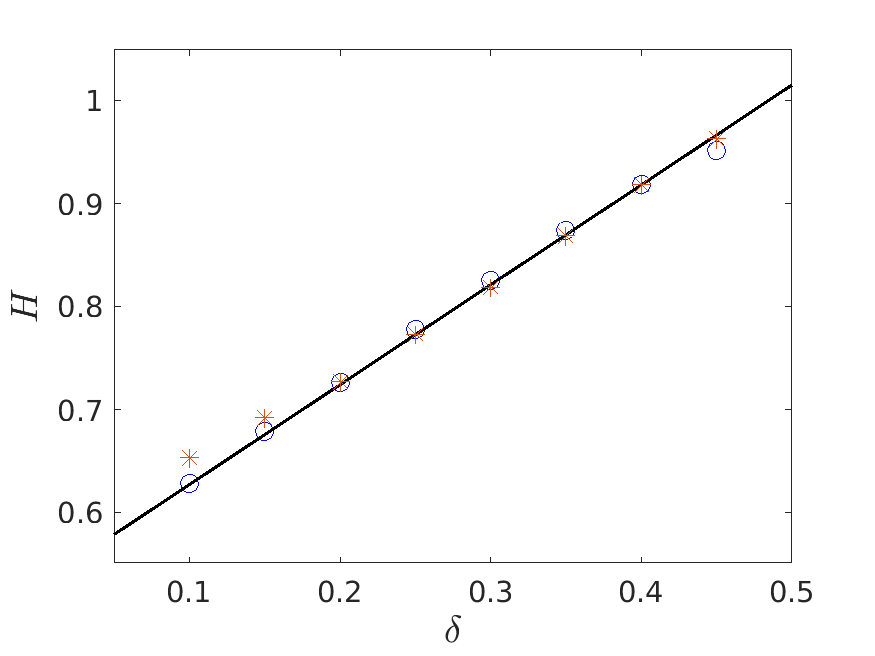}
  \caption{The plot of the Hurst exponent versus the parameter $\delta$.
  The data marked by circle and star are obtained from the stochastic process $z_x (t)$ and the characteristics area $a_c (L)$, 
  respectively. The straight guide line represents the relation between them, $H=0.94\delta + 0.5$. 
 }
\label{hdelta}
\end{figure}

In the Fig. \ref{cv}, the guide line represents $m=0.45$ for $\tau=1$, which indicates 
$H \approx 0.72$ that is different from that directly measured by the MSD of $z_x (t)$,
and the larger the time interval $\tau$, the smaller the value of the slope,
which is also different from the result of the NMRP giving the same values of the slope 
irregardless of the time interval $\tau$ at large time. 
It means that although single paths are separately taken, the water flows affected by previous flows in the neighboring
sites is not just mapped to a one-dimensional stochastic process, 
unlike the Scheidegger model without any memory for previous steps, and it is needed to be further understood
analytically.

Meanwhile, the values of exponents for real basins span a range rather than a specific value.
If the parameter $\delta$ is adjusted, different values of the exponent can be obtained. Figure \ref{hdelta}
shows the plot of $H$ obtained from the MSD of $z_x (t)$ and the characteristic area versus the parameter $\delta$.
The value of $H$ is almost linearly increased as $\delta$ increases, and especially for $0.2 < \delta < 0.3$, 
it matches the real values well. The slight difference between two values of $H$ 
for small $\delta$ can be originated from the initial behaviors that could be affected differently  according to 
the development of drainage area.

In conclusion, we have introduced a stochastic model that describes very well  the fractal characteristics of real river basins.
It suggests that river networks can be understood with only the viewpoint of stochastic process of water flow, 
which  have good points to describe the unified scaling laws of
real river basins in respect that a simple rule without consideration for the details of the contexts makes it be possible.
The key point in explaining the persistency in river flows is the idea of the dynamical replication probability 
from which although the stochastic process is Markovian, persistency can be generated. 

Such dynamical probability was firstly tried in the type of  walk models as far as we know, but there is another example 
where dynamical probability plays a key role in constructing a stochastic system. In the pioneering  paper \cite{cn} 
which opened the field of complex networks \cite{cnrev} describing the scale-free characteristics of a variety of  
complex systems 
such as economic, social, and biological systems as well as WWW and internet systems, the scale-free network is formed  
by the preferential attachment rule where the probability to attach an edge to a node in a complex network
depends on the number of edges of the node at each time and the larger the number of edges, the larger  the probability is.
In other words, a network is constructed by the probability determined by dynamical edges, meanwhile 
the network configuration with the distribution of edges is determined by the probability,
like this model where drainage areas determine the probability constructing a river network and in turn the probability 
determines the distribution of drainage areas.
Thus the dynamical probability could be an important origin for construction of complex systems which are developed
by a stochastic process affected by interactions between many components consisting of a system.

The concept of dynamical probability can be applied to other various systems such as economy, climate changes, 
analysis of big data, {\it etc} as well as fundamental science of which stochastic natures are in a fashion of 
the abnormal and yet little understood. And we expect it is able to deepen our understanding for the 
meaning and role of dynamical probability as well as for those systems.

\end{document}